\documentstyle [12pt]{article}

\begin{document}

\begin{flushright}
TCD--6--93 \\
September 1993
\end{flushright}

\vspace{8mm}

\begin{center}

{\Large\bf Singularities in Low Energy $d = 4$ Heterotic     \\
\vspace{3ex}
String and Brans-Dicke Theories  }  \\
\vspace{12mm}
{\large S. Kalyana Rama}

\vspace{4mm}
School of Mathematics, Trinity College, Dublin 2, Ireland. \\
\vspace{1ex}
email: kalyan@maths.tcd.ie   \\
\end{center}

\vspace{4mm}

\vspace{4mm}

\begin{quote}
ABSTRACT.  We present the static, spherically symmetric
solutions of the low energy $d = 4$
string. They form a simple generalisation of the
Schwarzschild solution and describe the gravitational
field of a spherical star in perturbative string theory.
If the Robertson parameter $\gamma$ is different from
one, these solutions describe singular objects with
naked singularities at their horizons, and whose Hawking
temperature is infinite. These
objects might represent ``massive remnants'', the
end states of black holes. Our analysis also implies that Brans-Dicke
theory has a naked singularity. Invoking cosmic censorship conjecture
will rule out Brans-Dicke theory.

\end{quote}

\newpage

The study of heterotic string
in four dimensions may lead to new physical phenomena and may
also reveal how intrinsically stringy effects come into play in
the four dimensional world at ordinary energies.
While a complete four dimensional string theoretic action
is not yet known, there is an effective action
at low energy that can be studied
to the lowest order in string tension and in string
coupling. This action always contains a graviton, a dilaton, and an
axion and leads to modifications
of general theory of relativity (GTR). Also, in the absence of
an axion, it is similar to Brans-Dicke (BD) theory, another
alternative to GTR.
BD theory is described by a parameter $\omega$,
related to the Robertson parameter
$\gamma$ by $\gamma = \frac{\omega + 1}{\omega + 2}$ \cite{sw},
and reduces to GTR in the limit $\omega \rightarrow \infty$.
Experimentally, $\gamma$ is measured to be $1 \pm .002$ \cite{viking}
and hence $\omega > 500$. Although the analysis presented here
also applies to BD type theories, we will concentrate on strings
in the following.

In GTR, the static, spherically symmetric
solution of Einstein's equations was given by Schwarzschild.
It describes the gravitational field of a spherical star
of mass $M$, which is a black hole if its radius
is less than the Schwarzschild radius. This black hole has
a singularity which is hidden behind a horizon.

The static, spherically symmetric solutions of the low energy
four dimensional
heterotic string would describe the gravitational field of
a spherical star modified by perturbative string effects.
In this letter we study them and explore their singularity structure.
Another motivation for this study comes from a desire to understand
better the recently discovered singular solution \cite{r}
in $d = 2$ string black hole \cite{rest}
where the curvature diverges at the horizon due to tachyon
back reaction effects.

Specifically, we look for a static spherically symmetric solution of
the $\beta$-function equations for graviton, dilaton, and
axion in the four dimensional critical heterotic string.
We find a two parameter family of
solutions where one parameter ($M$) is the mass of a star and
another ($b$) is a measure of radial variation of the dilaton $\phi$.
We analyse our solutions both in ``Einstein'' frame, which is commonly
used in literature, and in ``string'' frame, which describes
the metric seen by the propagating string.
Our results, which are also applicable to BD type theories,
are as follows:
(i) $b = 0$, {\em i.e.} \ $\phi = constant$,
gives the Schwarzschild solution of GTR, while
$b \neq 0$ gives a simple generalisation of the Schwarzschild solution.
(ii) The static solutions describe objects which have singular horizons
where the curvature invariants diverge. These singularities are naked.
(iii) The Hawking temperature of these singular objects,
defined to be proportional to the surface gravity at the horizon,
is infinite.

Just as the black holes, described by the static, spherically
symmetric solutions of GTR, are expected to form in
the collapse of massive stars, the singular objects
described by the above solutions
can also be naturally expected to form in such processes.
The implications of these results are as follows:
(i) String theory admits  solutions with naked
singularities thereby offering
an oppurtunity to study their formation. In this context one can
also study additional higher order stringy
interactions that may remove these singularities.
(ii) If $\gamma$ is measured to be different from one, then naked
singularities exist in the context of low energy perturbative
string theory and BD type theories.
(iii) The singular objects described here,
which have infinite Hawking temperature,
might represent ``massive remnants'' \cite{gidd},
the end states of black holes evaporating
by Hawking radiation.
(iv) If cosmic censorship conjecture (CCC)
is evoked to rule out the static, singular solutions, it will also
set $\gamma = 1$ and  $\omega = \infty$, thereby ruling out
Brans-Dicke theory. Evoking CCC would also necessitate imposing enough
(as yet unknown)
restrictions so that naked singularities will be absent
in any physical process.

We will consider the four dimensional low energy heterotic string with
graviton $(\tilde{g}_{\mu \nu})$, dilaton $(\phi)$, and axion ($\Theta$)
fields. In the sigma model approach the $\beta$-function equations
for these fields can be derived from an effective action
\begin{equation}\label{starget}
S = - \frac{1}{16 \pi G}
\int d^4 x \sqrt{\tilde{g}} \, e^{\phi} \,
( \tilde{R} - \tilde{a} (\tilde{\nabla} \phi)^2
+ \frac{1}{12} H_{\mu \nu \lambda} H^{\mu \nu \lambda}  + \Lambda)
\end{equation}
in the target space with coordinates
$ x^{\mu}, \; \mu = 0, 1, 2, 3$ where
$G$ is Newton's constant,
$H_{\mu \nu \lambda} = 3 \partial_{[\mu} B_{\nu \lambda]}$
is the field strength for the antisymmetric tensor $B_{\nu \lambda}$
and $\tilde{a} = 1$ for strings \cite{gm}.
Our choice of ``Riemann sign''
is $( \, - \, )$ in the notation of \cite{mtw}.
Note that the field $e^{- \frac{\phi}{2}}$ acts as a string coupling.
The axion $\Theta$ is related to $H_{\mu \nu \lambda}$
by the differential geometric relation
\[
H = - \frac{1}{2} e^{- 2 \phi} *d \Theta
\]
where $H$ is the three-form and $*$ is the Hodge dual.
The cosmological constant
$\Lambda = 0$ for a critical string compactified to four dimensions
and $\frac{d - 10}{2}$ for a non critical string in $d$ dimensions.
Brans-Dicke type theories are described by (\ref{starget})
if $\Theta = 0$ and $\tilde{a} < 0$ \cite{mtw}.
In the effective action above, which is written in ``string''
frame with metric $\tilde{g}_{\mu \nu}$, the curvature term is
not in the standard Einstein
form. However, the standard form
can be obtained by a dilaton dependent conformal transformation
\begin{equation}\label{conf}
\tilde{g}_{\mu \nu} = e^{- \phi} g_{\mu \nu}
\end{equation}
to the ``Einstein'' frame with metric $g_{\mu \nu}$.
The effective action then becomes \cite{gm}
\begin{equation}\label{etarget}
S = - \frac{1}{16 \pi G} \int d^4 x \sqrt{g} \,
( R + \frac{a}{2} (\nabla \phi)^2
+ \frac{1}{2} e^{- 2 \phi} (\nabla \Theta)^2
+ e^{- \phi} \Lambda)
\end{equation}
where $a = 3 - 2 \tilde{a} \, (\, = 1$ for strings and $> 3$ for
BD theory).
The equations of motion for
$g_{\mu \nu}, \, \phi$ and $\Theta$ that follow from this action
are
\begin{eqnarray}\label{beta}
2 R_{\mu \nu} + a \nabla_{\mu} \phi \nabla_{\nu} \phi
+ e^{- 2 \phi} \nabla_{\mu} \Theta \nabla_{\nu} \Theta
+ g_{\mu \nu} e^{- \phi} \Lambda  & = & 0 \nonumber \\
a \nabla^2 \phi + e^{- 2 \phi} (\nabla \Theta)^2
+ e^{- \phi} \Lambda & = & 0 \nonumber \\
\nabla ( e^{- 2 \phi} \nabla \Theta ) & = & 0  \; .
\end{eqnarray}
In the following we will consider the critical string, and hence
$\Lambda = 0$.
We will look for the static, spherically symmetric solutions
of these equations in the Schwarzschild gauge where
$d s^2 = - f d t^2 + f^{- 1} d r^2 + h^2 d \Omega^2, \;
d \Omega^2$ being the line element on an unit sphere.

Writing $f = e^{\sigma}$, equations (\ref{beta})
can be written in a simple form as
\begin{eqnarray}\label{babe}
\sigma'' + \sigma' ( \frac{f'}{f} + \frac{2 h'}{h} ) & = & 0
\nonumber \\
\frac{4 h''}{h} + a \phi'^2 + e^{- 2 \phi} \Theta'^2
& = & 0 \nonumber \\
(f h^2)'' - 2 & = & 0 \nonumber \\
a \phi'' + a \phi' ( \frac{f'}{f} + \frac{2 h'}{h} )
+ e^{- 2 \phi} \Theta'^2 & = & 0 \nonumber \\
\Theta'' + \Theta' ( \frac{f'}{f} + \frac{2 h'}{h} )
- 2 \phi' \Theta' & = & 0
\end{eqnarray}
where $'$ denotes $r$-derivatives. Integrating the
first and third equations above gives
$\sigma' f h^2 = k r_0$ and $f h^2 = r (r - r_0)$
where  $k$ and $r_0$ are (real) integration constants.
The $\sigma'$ equation
can again be integrated using the expression for $f h^2$ to give
\begin{eqnarray}\label{fh}
f & = & (1 - \frac{r_0}{r})^k \nonumber \\
h^2 & = & r^2 (1 - \frac{r_0}{r})^{1 - k}  \; .
\end{eqnarray}
We consider only the case $k > 0$ in the following.
In deriving (\ref{fh}) we have shifted $r$ and scaled $f$
appropriately \cite{scale}. Requiring $f$ to describe
the gravitational field produced by a mass $M$ in the
asymptotic region ($r \rightarrow \infty$) gives
$r_0 = \frac{2 M}{k}$. The curvature scalar is
\begin{equation}\label{cur}
R = - \frac{r_0^2 (1 - k^2)}{2 r^4 (1 - \frac{r_0}{r})^{2 - k}}  \; .
\end{equation}
The quadratic curvature invariants are
$R_{\mu \nu} R^{\mu \nu} \, ( \, = R^2$ in the Schwarzschild gauge),
and the Gauss-Bonnet combination
${\cal G} = R_{\mu \nu \lambda \rho}  R^{\mu \nu \lambda \rho}
- 4 R_{\mu \nu} R^{\mu \nu} + R^2$ given, for the above solution, by
\[
{\cal G} = \frac{4 r_0^2 k}
{r^6 (1 - \frac{r_0}{r})^{4 - 2 k}}
\{ 3 k (1 - \frac{r_0}{r})^2
+ k (1 - k) \frac{r_0}{r} (1 - \frac{r_0}{r})
- \frac{(1 - k)^2 (2 - k) r_0^2}{4 r^2} \} \; .
\]
The curvature invariants are singular at $r = 0$ for any $k$
and, now, are also singular at
$r = r_0$ if $k \neq 1$ and $< 2$. These singularities are
naked.

Consider the case when only a dilaton or an axion is present.
Let $\Theta = 0$. Then $\phi''$-equation in (\ref{babe}) gives
$\phi' f h^2 = b r_0$ and, hence
\begin{equation}\label{phi}
e^{\phi} = e^{\phi_0} (1 - \frac{r_0}{r})^b
\end{equation}
where
$\phi_0$ is the value of dilaton at $r = \infty$
and $b$ is a constant which, if positive, corresponds to strong
coupling and, if negative, to weak coupling at the horizon.
Note that replacing $\phi$ by $\Theta$ will give the solution
when only an axion is present.
Substituting for $\phi'$ in $h''$-equation in (\ref{babe}) gives
\begin{equation}\label{k}
k = \sqrt{1 - a b^2} \; .
\end{equation}
Hence $k \leq 1$ and the equality holds only when $b = 0$.
Also, $a b^2 < 1$ since $k$ is real and strictly positive.
Since $a \geq 1$ this implies that $|b| < 1$.
Thus equations (\ref{fh})-(\ref{k})
form the static, spherically symmetric solutions
of equations (\ref{beta}) in the
Schwarzschild gauge when only a dilaton or an axion is
present. We will later comment
on the solutions when both are present.

The above Einstein frame solution can be written in string
frame \cite{schwein} as
\begin{eqnarray}\label{sfh}
\tilde{f} & = & e^{- \phi_0} (1 - \frac{r_0}{r})^{k - b} \nonumber \\
\tilde{h}^2 & = & e^{- \phi_0} r^2
(1 - \frac{r_0}{r})^{1 - k - b} \nonumber \\
\tilde{R} & = & \frac{r_0^2 (1 - k^2) e^{\phi_0}}
{r^4 (1 - \frac{r_0}{r})^{2 - k - b}}
\end{eqnarray}
where $\tilde{\;}$s refer to string frame and
$\tilde{r}$ is related to $r$ by
\begin{equation}\label{rr}
\tilde{r} + constant = e^{- \phi_0}
\int d r (1 - \frac{r_0}{r})^{- b} \; .
\end{equation}
We will set $\phi_0 = 0$ so that $\tilde{f}$ is one
asymptotically, and further define the mass $\tilde{M}$ in
string frame by $r_0 = \frac{2 \tilde{M}}{k - b}$. Since $|b| < 1$,
the integral in (\ref{rr}) is not divergent near the
horizon. Hence $\tilde{r}$ and $\tilde{r}_0$, the horizon
in string frame, are well defined.
Thus the curvature invariants in the string frame are also
divergent at the horizon if $k \neq 1$ and $k + b < 2$ and these
singularities are also naked in the string frame.

The Robertson parameters, which can be measured by experiments,
are obtained by writing the line element in isometric form and
expanding asymptotically the various functions in it \cite{sw}.
In our case they are given by $\alpha = \beta = 1$, and
$\gamma = 1 $ in Einstein frame, and
$\gamma = \frac{k + b}{k - b}$ in string frame \cite{rob}.
The measured value of $\gamma$ is $1 \pm .002$ \cite{viking}.

Let $A(r)$ be the area of a sphere $S_r$ of radius $r$ centered at
$r = 0$ and $\kappa = \frac{f'}{2}$ be the gravitational field
in Einstein frame. Let $\tilde{A}(r)$ and
$\tilde{\kappa} = \frac{1}{2} \frac{d \tilde{f}}{d \tilde{r}}$
denote the corresponding quantities in string frame.
We then have
\[
A(r) = 4 \pi r^2 (1 - \frac{r_0}{r})^{1 - k}  \; , \; \;
\tilde{A}(r) = 4 \pi r^2 (1 - \frac{r_0}{r})^{1 - k - b}
\]
and
\[
\kappa (r) = \frac{M}{r^2
(1 - \frac{r_0}{r})^{1 - k} }  \; , \; \;
\tilde{\kappa} (r) =
\frac{\tilde{M}}{r^2 (1 - \frac{r_0}{r})^{1 - k} }  \; .
\]

We will now discuss the properties of the above solutions
for $k > 0$.

(i) $b = 0$, {\em i.e.} \ $k = 1$, gives the standard
Schwarzschild solution.

(ii) There is only one horizon, at $r = r_0$ (where $f = 0$).

(iii) The curvature invariants
diverge at $r = 0$ for any $k$. Now they also diverge at
the horizon if $b \neq 0$. These singularities are naked.

\underline{Einstein Frame} :

(iv) The area of the horizon shrinks to zero if $b \neq 0$.

(v) The gravitational field $\kappa$
obeys a ``Gauss law''
$\int_{S_r} \, \kappa(r)  d A = 4 \pi M$.

(vi) Taking the Hawking temperature $T$
and the black hole entropy $\Sigma$ to be given respectively by
$T = \frac{\kappa(r_0)}{2 \pi} $
and $\Sigma = \frac{A(r_0)}{4}$ we have,
if $b \neq 0$,
\[
T = \infty \; , \; \; \;     \Sigma = 0 \; , \; \;
T \Sigma  = \frac{M}{2}  \; .
\]

\underline{String Frame} :

The following properties can be easily seen using a graph of
$k \pm b$ vs $b$ \cite{kpmb}.

(iv) The area of the horizon diverges if $b > 0$ and
shrinks to zero if $b < 0$.

(v) The ``Gauss law'' in this case is
$\int_{S_r} \, \tilde{\kappa}(r)  d A
= 4 \pi \tilde{M} (1 - \frac{r_0}{r})^{- b}$.

(vi) $T = \infty$ if $b \neq 0 \, . \; \Sigma$ and $T \Sigma$ are both
infinite if $b > 0$ and zero if $b < 0$.

We emphasise the point that the static solutions given here
describe objects which exhibit naked singularities
and have infinite Hawking temperature unless $b = 0$, that is
$\gamma = 1 , \, \phi = constant$. We also note that
the Hawking temperature $T$ is defined here to be proportional
to the surface gravity at the horizon and that
it is important to find out by a quantum field theoretic calculation
whether these singular objects
indeed emit Hawking radiation at temperature $T$.

We now consider equations (\ref{babe}) with both dilaton and axion
present. With $\chi' \equiv e^{- \phi} \Theta'$ we get
$\chi'' + \chi' (\frac{f'}{f} + \frac{2 h'}{h})
- \chi' \phi' = 0$.
Multiplying this
equation by $\chi'$ and the $\phi''$-equation by $\phi'$ and
adding them together gives
\[
f h^2 \sqrt{a \phi'^2 + \chi'^2} = B r_0 \sqrt{a}
\]
where $B$ is a constant. From this and the $h''$-equation it follows
that
\[
k = \sqrt{1 - a B^2}
\]
and, the previous analysis can be carried
through in Einstein frame with $b$ replaced by $B$.
In particular, the singularities at the horizon
will be absent if and only if $B = 0$, which implies that both dilaton
and axion fields are constants. Thus, including the axion does not
remove the singularities at the horizon.

The above static solutions
describe the gravitational field outside a spherical star of
mass $M$ in string and BD theories. If the radius of the star is
less than $r_0$ it will exhibit naked singularities
at its horizon in both Einstein and string frame unless $b = 0$,
{\em i.e.} \ $\gamma = 1, \, \phi = constant$.

Can such stars form? In GTR the static, spherically symmetric \\
Schwarzschild solution
describes a black hole if the radius of the star is less
than the Schwarzschild radius. Such black holes
are believed to form as massive stars collapse.
In a similar way, it is natural to expect that in string and
BD theories massive stars will collapse to singular objects, described
by the static, spherically symmetric
solutions in these theories, which exhibit naked
singularities. However, more analysis is needed for
a definitive answer.

At this stage there are two alternatives to take.
The first is to evoke cosmic censorship
conjecture and rule out such singular solutions. This would imply
that $\gamma = 1$, and hence $\omega = \infty$,
thus ruling out Brans-Dicke theory. But
naked singularities may appear
in some dynamical process described by these theories. For
example, such a process appears feasible
in $d = 2$ string, resulting in the formation of
singular objects \cite{scene}.
Hence one also needs to ensure the absence of singularities in any
dynamical process. Furthermore, if
$\gamma$ were experimentally found to be different from one it would
imply, within the context of string and BD theories,
the existence of naked singularities. Therefore evoking
CCC does not seem fruitful.

The second alternative is to consider the possibility that such
singular solutions exist. This would mean that we have to understand
the singularities and find ways of dealing with all the
conundrums associated with them. It is also
possible that in string theory (although not in BD theory)
these singularities will be removed by higher order corrections
or by some other mechanism which is intrinsically stringy.
Moreover, physically, such singular solutions could describe
``massive remnants'' \cite{gidd}, the
end states of black holes evaporating by Hawking radiation.

At the very least, low energy string (and possibly BD) theory
gives us a model where naked singularities are likely to evolve.
This would provide an oppurtunity to study their actual
evolution and ways of understanding them.
Also, taking the low energy string model
in its own right, one might postulate or derive from a more
fundamental theory additional interactions,
symmetries, etc.\ that
would ensure the absence of naked singularities. If naked
singularities are indeed absent in nature, one outcome of this
work would be to rule out Brans-Dicke theory and its attendent
successes. However, one might invoke a time dependent $\omega$ that
relaxes to $\infty$ in the end, thereby avoiding the problem of
the static, naked singularity. But, in such a theory,
one needs to ensure the absence of naked singularities during the
evolution of $\omega$ and also during any other physical process.

Besides the issues related to naked singularities, one might
look for the effects of the static solution presented here
in other contexts, as for example, in
stellar dynamics where strong gravitational
fields are involved and where the deviations from GTR
may have considerable effects. It would be very interesting if
the consequences of the solution presented here differ from those of
the Schwarzschild solution in any appreciable and detectable way.

\vspace{4ex}

It is a pleasure to thank S. Ominsky for reading the manuscript,
M. J. Perry for a helpful correspondence, and S. Sen for encouragement.
This work is supported by EOLAS Scientific Research Program
SC/92/206.

\vspace{4ex}

{\bf Note Added}:
After the completion of our work, we were informed of
\cite{fucito}, where instability of stringy axion black hole was studied,
and \cite{lousto}, where BD theory was analysed in a way similar to ours.


\begin{thebibliography}{999}
\bibitem{scale}
Various fields can be scaled/shifted by constant factors/terms
using the invariance of equations (\ref{babe}) under
$ r \rightarrow a r + a_0 , \;
\phi \rightarrow \phi + \phi_0 , \;
h \rightarrow b h , \;
f \rightarrow \frac{a^2}{b^2} f, \; $ and
$\Theta  \rightarrow e^{2 \phi_0} \Theta$.
\bibitem{schwein}
In the string frame, the line element in the Schwarzschild gauge
is given by
$d \tilde{s}^2 = - \tilde{f} d t^2 + \tilde{f}^{- 1} d \tilde{r}^2
+ \tilde{h}^2 d \Omega^2 $
where $\tilde{f} = e^{- \phi} f , \;
\tilde{h} = e^{- \frac{\phi}{2}} h , \; $ and
$d \tilde{r} = e^{- \phi} d r$.
\bibitem{rob}
In Einstein frame, the isotropic form is given by \\
$d s^2 = - f d t^2 + F (d \rho^2 + \rho^2 d \Omega^2)$ where
$f = (1 - \frac{r_0}{4 \rho})^{2 k}
(1 + \frac{r_0}{4 \rho})^{- 2 k} $ and \\
$F = (1 + \frac{r_0}{4 \rho})^{2 (1 + k)}
(1 - \frac{r_0}{4 \rho})^{2 (1 - k)}$.
The corresponding functions in the string frame are
$\tilde{f} = (1 - \frac{r_0}{4 \rho})^{2 (k - b)}
(1 + \frac{r_0}{4 \rho})^{- 2 (k - b)}$ and \\
$\tilde{F} = (1 + \frac{r_0}{4 \rho})^{2 (1 + k + b)}
(1 - \frac{r_0}{4 \rho})^{2 (1 - k - b)}$.
These functions can be expanded near $\rho = \infty$ to obtain
the Robertson parameters.
\bibitem{kpmb}
The graph of $(k - b)$ can be described by giving
its values at special points of $b$ lying in the allowed range
$- \frac{1}{\sqrt{a}} \leq b \leq \frac{1}{\sqrt{a}}$ as follows:
\[
\begin{array}{cccccccc}
(k - b) & : & (\frac{1}{\sqrt{a}}, \, &  1, \,&  \sqrt{1 + \frac{1}{a}},
\, & 1, \, & 0, \, &  \frac{- 1}{\sqrt{a}} ) \\
b & : &
( \frac{- 1}{\sqrt{a}}, \, & \frac{- 2}{a + 1}, \, &
\frac{- 1}{\sqrt{a (a + 1)}}, \, &  0, \, &  \frac{1}{\sqrt{a + 1}}, \, &
\frac{1}{\sqrt{a}} )
\end{array} \; .
\]
The variation of $(k - b)$ is smooth in between the above points.
Note that $(k - b)_{max} = \sqrt{1 + \frac{1}{a}} \leq \sqrt{2}$
since $a \geq 1$ and that the behaviour of $(k + b)$ follows
by changing $b$ to $- b$.
\end{thebibliography}
\end{document}